\documentclass[aps,pre,onecolumn,notitlepage]{revtex4-2}

\usepackage[margin=2.4cm]{geometry}
\usepackage{amsmath,amssymb,amsthm}
\usepackage{mathtools}
\usepackage{bm}
\usepackage{physics}
\usepackage{enumitem}
\usepackage{hyperref}
\usepackage{subcaption}
\usepackage{float}
\usepackage{xcolor}
\usepackage{tikz}

\begin{document}

\title{Complete local expansion of the availability function in random sequential adsorption
of aligned squares at low density: Termination at fourth order}
\author{F.\ \c{T}olea}
\affiliation{National Institute of Materials Physics, POB MG-7, Bucharest-M\u{a}gurele, 077125, Romania}

\author{M.\ \c{T}olea}
\email{tzolea@infim.ro; tzolea123@yahoo.com}
\affiliation{National Institute of Materials Physics, POB MG-7, Bucharest-M\u{a}gurele, 077125, Romania}

\begin{abstract}
We consider random sequential adsorption (RSA) of aligned squares and derive the low-coverage expansion of the availability function $\alpha(q)$, the fraction of positions accessible to an additional square, up to fourth order in the coverage $q$. At low coverage, the reduction of available space can be understood in terms of geometric overlap between exclusion regions created by previously deposited squares. A single square blocks a finite area; pairs of squares may have overlapping exclusion zones, reducing the total blocked area; similarly, three and four squares can share a common overlap region, leading to higher-order corrections. These contributions can be systematically accounted for through an inclusion--exclusion expansion based on the geometry of overlapping exclusion regions, with alternating signs dictated by inclusion--exclusion. The expansion terminates exactly at fourth order, since no more than four deposited squares can simultaneously overlap the exclusion region of a trial insertion. The coefficients are obtained by explicit enumeration of all such geometrically admissible configurations and are further confirmed by numerical simulations on a discrete lattice, showing agreement within statistical uncertainty.
\end{abstract}

\maketitle

\section{Introduction}

Let us begin with the motivation for the present work, even though the problem we address is not exactly the same. The average coverage of a line by randomly placed, non-overlapping identical segments admits an elegant solution first obtained by R\'enyi~\cite{Renyi1958}, who showed that the asymptotic filling fraction approaches a universal constant $\phi_1 = 0.7476...$ (a brief derivation is provided in Appendix~A for completeness).
The one-dimensional problem has been extensively studied and admits several complementary formulations, including exact integral representations, asymptotic analyses, and numerical approaches~\cite{Renyi1958,Dvoretzky1964,Weiner1969,Marsaglia1989,Krapivsky1992,Slavik2023}. These developments highlight that even in one dimension, the process involves subtle correlations arising from excluded regions and sequential constraints. On the other hand, this one-dimensional result is not only classic, but also deceptive: it suggests that closely related higher-dimensional problems might admit similarly simple answers.

Indeed, once one passes from segments on a line to extended objects on a surface, the geometry becomes far less forgiving. In 1960, Pal\'asti conjectured that the average filling fraction in dimension $n$ is given by a simple power of the one-dimensional result~\cite{Palasti1960}, namely $\phi_n=\phi_1^n$. In particular, for the two-dimensional case of randomly deposited aligned squares, this predicts $\phi_2=\phi_1^2=0.5589...$ Early numerical studies did find values close to $\phi_2\approx0.5590...$~\cite{Mackenzie1962}, apparently consistent with Pal\'asti's prediction within their uncertainty. Yet more precise simulations later established a slightly larger value, $\phi_2\approx0.5620...$~\cite{PrivmanWangNielaba1991,Dickman1991}. The conjecture thus turned out to be remarkably accurate, but not exact.

Beyond this specific historical conjecture, the random filling of a surface by non-overlapping objects is of central importance in physics, where it is known as random sequential adsorption (RSA) (see, e.g., Refs.~\cite{PrivmanWangNielaba1991, Dickman1991, Feder1980, Evans1993, Cadilhe2007, Talbot2000, Pomeau1980, Swendsen1981,Viot1992,Bartelt1992,Meakin1992,Torquato2006,Zhang2013,Ciesla2021,Lebovka2023}). In this process, particles are deposited one by one at random positions, and any attempted placement that overlaps previously adsorbed particles is rejected, making the dynamics irreversible and leading to a disordered jammed state in which no further particle can be inserted. As a minimal and widely used model for irreversible deposition, RSA has found applications ranging from the adsorption of atoms and molecules on solid surfaces~\cite{PrivmanWangNielaba1991,Evans1993}, to colloidal particle deposition~\cite{Feder1980}, thin-film growth~\cite{Evans1993}, and biological systems such as protein adsorption on membranes~\cite{Talbot2000}. The kinetics of the approach to jamming, characterized by slow relaxation and the buildup of spatial correlations, have been extensively studied~\cite{Pomeau1980,Swendsen1981,Krapivsky1992}. While the one-dimensional case is exactly solvable and leads to R\'enyi's constant~\cite{Renyi1958}, higher-dimensional systems exhibit complex geometric correlations and jamming phenomena that remain analytically challenging~\cite{Evans1993,Cadilhe2007,Torquato2002}.

A key quantity of interest is the jamming coverage, defined as the maximal achievable filling fraction under irreversible sequential deposition. Unlike equilibrium packing problems, RSA systems are intrinsically out of equilibrium, and their final state depends on the history of the adsorption process. This history dependence generates nontrivial spatial correlations and a slow approach to saturation, features that become increasingly pronounced in higher dimensions~\cite{Evans1993,PrivmanBook1991}.

For this reason, the low-density regime is especially valuable. It is the regime in which one can still hope to extract exact information before the full complexity of the global deposition history takes over. At low coverage, RSA can be described by systematic expansions of the adsorption probability, or available-surface function (ASF), which plays a role analogous to a virial expansion in equilibrium statistical mechanics~\cite{Evans1993,Talbot2000,MayerMayer1940}. One writes $\alpha(q)=1-C_1q+C_2q^2-C_3q^3+\cdots$, where the coefficients $C_n$ depend on particle shape and orientation.

For circular disks in two dimensions, numerical and series-expansion studies yield approximate values $C_2\approx3.3$ and $C_3\approx4.7$--$4.8$~\cite{Dickman1991}. For anisotropic particles, second- and third-order coefficients have been obtained numerically~\cite{Ricci1992}. However, explicit closed-form expressions derived directly from geometric overlap configurations are generally not available beyond leading order. This is the gap addressed here for the case of aligned squares, where we derive the local low-density expansion explicitly up to its maximal geometrically allowed order.

In the present paper, we focus on the continuum RSA of aligned squares at low occupancy and in particular study the availability function $\alpha(q)$, which gives the fraction of positions where one additional square can still be inserted at coverage $q$. Our strategy is elementary in spirit but complete within its domain. Rather than approximating the coefficients numerically or inferring them indirectly, we construct the low-density expansion directly from geometric overlap clusters. The logic is progressive: first one accounts for the exclusion generated by a single square, then one corrects the overcounting produced when two exclusion zones overlap, then one corrects the overcorrection produced by triple overlaps, and so on. In this way, each order in the expansion refines the previous one by going one step further in the overlap hierarchy.

In this framework, inclusion--exclusion determines the alternating structure of the expansion, while the magnitude of each contribution is fixed by the geometry of overlap regions, which can be expressed compactly through suitable overlap kernels.

A central observation is that, for aligned squares, no more than four mutually non-overlapping deposited squares can simultaneously overlap the exclusion region of a trial square. This means that the fourth-order term is not merely the next correction in a formal series: it is the highest geometrically relevant local contribution. In that sense, the polynomial derived here is not an arbitrary truncation but the complete local overlap hierarchy permitted by the geometry.

More generally, this construction may be viewed as an explicit realization of the cluster-expansion viewpoint for RSA, analogous in spirit to virial or Mayer expansions in equilibrium statistical mechanics, but adapted to an irreversible process~\cite{Evans1993,Torquato2002}. To the best of our knowledge, the corresponding coefficients for the aligned-square availability function, up to this geometrically defined maximal local order, have not been written explicitly in closed form. Such coefficients may also serve as analytical benchmarks for numerical RSA simulations and for approximate descriptions of irreversible adsorption, deposition, and surface-coverage processes.

The remainder of this paper is organized as follows. In Sec.~II we present the geometric framework and derive the complete low-density expansion of the availability function, including all coefficients up to fourth order. To make the logic easy to follow, the derivation is organized step by step: at each order we indicate which part of the previous counting is being corrected and why the new term has its particular sign. Section~III is devoted to a discussion of the results and their implications. Finally, Sec.~IV summarizes the main conclusions. For completeness, Appendix~A recalls the classical one-dimensional result of R\'enyi, while Appendix~B provides the discrete lattice formulation and numerical validation.

\section{Low-density polynomial derivation of the availability function \texorpdfstring{$\alpha(q)$}{alpha(q)}}

Within the framework of RSA, squares are deposited randomly on a surface, one at a time, with the constraint that no overlaps are allowed. Each deposited square reduces the set of positions available for subsequent insertions, so the availability function naturally decreases as the coverage increases. The reduction of availability due to a single square---its associated exclusion zone---is straightforward to compute. However, when two or more squares are present, their exclusion zones may overlap, and the corresponding corrections to the available area become increasingly complex. The aim of the present work is to determine these corrections systematically.

The guiding idea of this section is simple. We begin with the exclusion created by one deposited square. This gives the first-order term. We then ask what was counted incorrectly at that level: the answer is that overlap regions of two exclusion zones were subtracted twice. Correcting that produces the second-order term. Repeating the same logic, one finds that triple-overlap regions are then overcorrected and must be subtracted, and finally quadruple-overlap regions must be added back. Thus the whole derivation follows a clear inclusion--exclusion story, order by order.

\subsection{Definition of the availability function}

Let $L$ be the side of the aligned squares and $\rho$ the number density of already deposited squares.
We define the dimensionless occupation variable
\begin{equation}
q = \rho L^2,
\label{eq:q_def}
\end{equation}
which represents the fraction of area covered by deposited squares.

\medskip

The central quantity of this work is the \emph{availability function} $\alpha(q)$, defined as the probability that a new trial square, placed uniformly at random in the plane, can be inserted without overlapping any previously deposited square. Equivalently, $\alpha(q)$ is the fraction of the plane where such a placement is possible.

\medskip

At a given stage of the process, the plane is partitioned into two complementary regions:
\begin{itemize}[leftmargin=*,itemsep=2pt]
\item the \emph{forbidden region}, where placing a new square would lead to overlap with existing ones,
\item the \emph{available region}, where a new square can be inserted.
\end{itemize}

Thus, $\alpha(q)$ is simply the fraction of the plane occupied by the available region.

\medskip

To describe the forbidden region geometrically, we associate to each deposited square its \emph{exclusion zone}, defined as the set of positions of the trial square that would produce overlap.

For aligned squares of side $L$, this exclusion zone is itself a square of side $2L$ (see Fig.\ref{fig:one_square}), with area $v_1 = (2L)^2 = 4L^2$. This is the geometric origin of the first coefficient: one deposited square blocks a square region with double side, so the corresponding first-order contribution must reduce the availability.

The forbidden region is therefore given by the union of all exclusion zones, and the availability function can be written as
\begin{equation}
\alpha(q) = 1 - \frac{\text{area of the union of exclusion zones}}{\text{total area}}.
\label{eq:alpha_union}
\end{equation}

\medskip

At low occupation, overlaps between exclusion zones remain local, and $\alpha(q)$ naturally admits a systematic expansion in powers of $q$.
Within the local cluster approximation (i.e., retaining only geometrically allowed overlap configurations up to four-body intersections), we seek a polynomial representation of the form
\begin{equation}
\alpha(q) = 1 + a_1 q + a_2 q^2 + a_3 q^3 + a_4 q^4.
\label{eq:alpha_ansatz}
\end{equation}

The coefficients $a_n$ encode the contributions of $n$-body overlaps between exclusion zones and will be derived explicitly in the following subsections.

\medskip

\noindent
\textit{Technical remark.} The position of a trial square may be parameterized by any reference point (e.g., its lower-left corner as we do here, see Fig.\ref{fig:one_square}, or its center, etc). This choice is purely conventional and does not affect $\alpha(q)$, as all such parameterizations are equivalent up to a rigid translation.

\subsection{Inclusion--exclusion structure of the expansion}

Before entering the explicit calculations, it is useful to state the overall logic once in abstract form. The first-order term treats each exclusion zone independently. That is correct as long as exclusion zones do not overlap. The moment two exclusion zones overlap, however, the common part has been removed twice. This is the first overcounting that must be repaired. Once that repair is made, regions common to three exclusion zones become overrepaired, and so on. The alternating signs of the coefficients are therefore not mysterious: they simply reflect the successive corrections required by inclusion--exclusion.

The expansion of the availability $\alpha(q)$ can be understood most naturally in terms of exclusion zones. We therefore briefly describe how the geometric structure of these regions directly determines the form and the signs of the expansion coefficients.

\begin{figure}[htbp]
  \centering
  \includegraphics[width=0.2\linewidth]{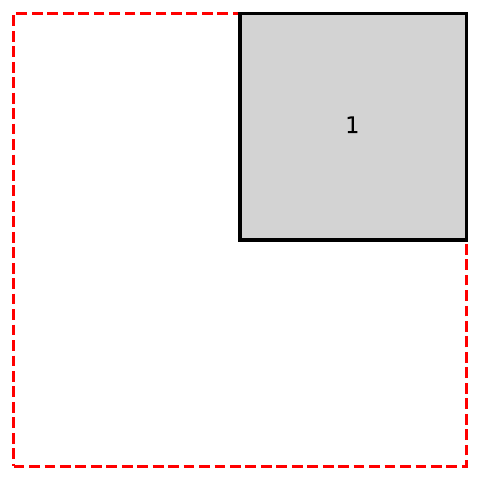}
  \caption{Exclusion zone associated with a deposited square "1" of side $L$.
The exclusion zone includes the square itself and forms a larger square of side $2L$.
The dashed square represents all positions of the lower-left corner of a second square that would produce overlap with the first one.
Any placement inside this region is therefore forbidden.}
  \label{fig:one_square}
\end{figure}

For a single deposited square, the exclusion zone is simply a square of side $2L$ (see Fig.~\ref{fig:one_square}). Its area represents the total forbidden region generated by one square. This contribution reduces the available area and therefore enters the expansion of $\alpha(q)$ with a negative sign, giving rise to the first-order term.

\begin{figure}[H]
\centering

\begin{subfigure}{0.25\textwidth}
    \centering
    \includegraphics[width=\linewidth]{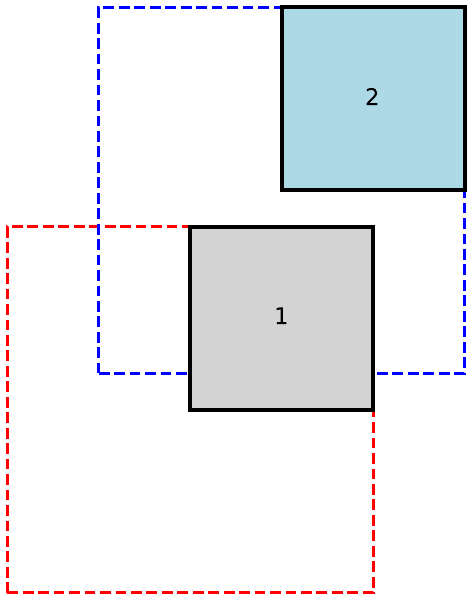}
    \caption{}
\end{subfigure}
\hspace{0.02\textwidth}
\begin{subfigure}{0.25\textwidth}
    \centering
    \includegraphics[width=\linewidth]{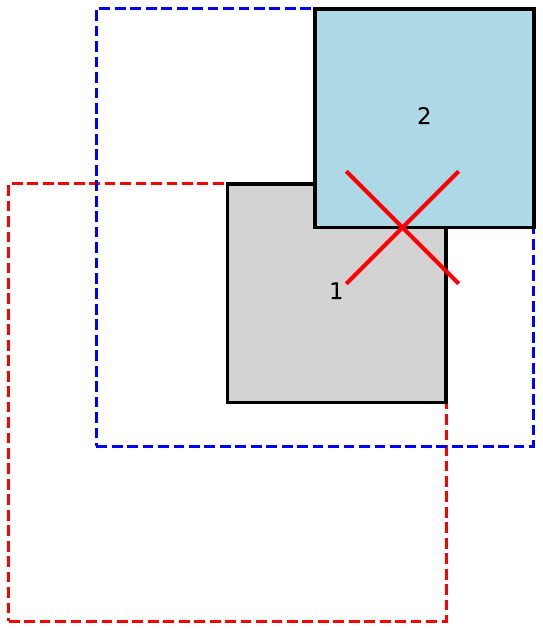}
    \caption{}
\end{subfigure}

\caption{(a) Two exclusion zones can overlap; the overlapping area must be added back to the availability, as it has been subtracted twice at first order (single-square contribution).
(b) The squares themselves cannot overlap.}
\label{Two_Squares}
\end{figure}

When two deposited squares are present, their exclusion zones may overlap (see Fig.\ref{Two_Squares}). In the naive sum of forbidden regions, this common part is counted twice. To correct for this overcounting, one must add back the overlap of the two exclusion zones. This produces a positive second-order contribution, which we shall calculate with care.

\begin{figure}[htbp]
  \centering
  \includegraphics[width=0.3\linewidth]{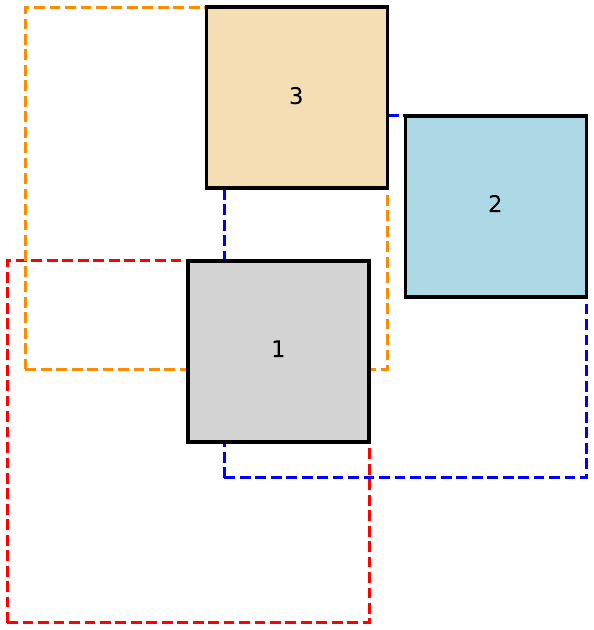}
 \caption{Example of overlap between three exclusion zones. The common region corresponds to the third-order correction.}
  \label{fig:three_squares}
\end{figure}

For three deposited squares, the same reasoning applies once more (see Fig.\ref{fig:three_squares}). After subtracting one-body contributions and adding all pairwise corrections, any region where three exclusion zones overlap has been overcorrected and must therefore be subtracted again. This yields a negative third-order contribution.

\begin{figure}[htbp]
  \centering
  \includegraphics[width=0.4\linewidth]{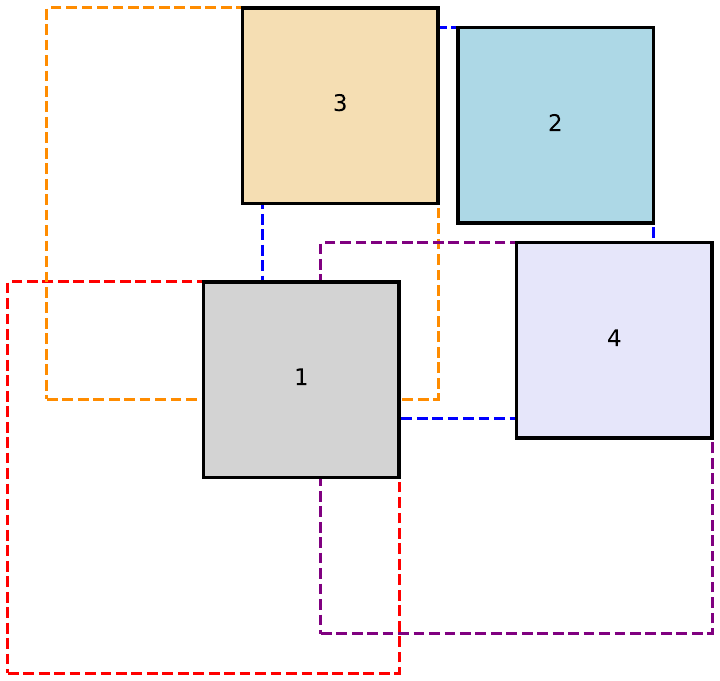}
\caption{Example of overlap between four exclusion zones. The common region corresponds to the fourth-order correction.}
  \label{fig:four_squares}
\end{figure}

The same alternating logic continues at order four (see Fig.\ref{fig:four_squares}) and is simply the inclusion--exclusion principle applied to the union of exclusion zones. In this framework, the coefficients of the expansion are determined entirely by the geometry of multiple overlaps of exclusion regions.

In particular, referring to Eq.~(\ref{eq:alpha_ansatz}), one immediately obtains the sign structure
\[
a_1 < 0,\qquad a_2 > 0,\qquad a_3 < 0,\qquad a_4 > 0,
\]
consistent with the alternating nature of inclusion--exclusion.

It is worth emphasizing that this alternation reappears inside the explicit coefficient calculations, as shall be detailed below. The allowed-domain integrals are obtained by starting from an unconstrained overlap integral and then applying inclusion--exclusion to the forbidden pair-overlap conditions between the deposited squares.

Thus, each coefficient $a_n$ corresponds to an $n$-body geometric overlap of exclusion zones, establishing a direct correspondence between the algebraic expansion and the underlying spatial structure of the problem.

\subsection{First-order coefficient \texorpdfstring{$a_1$}{a1}}

We start with the simplest level of description: each deposited square is considered separately, without yet worrying about overlaps between different exclusion zones. At this order there is nothing to correct, because only one exclusion zone is involved. The result is therefore immediate and serves as the baseline that all higher orders will refine.

As seen in Fig.~\ref{fig:one_square}, each deposited square of side $L$ ``takes away'' a surrounding area of $4L^2$, where no other square can be deposited without overlapping.
This leads to a total blocked fraction of
\begin{equation}
\rho\, 4L^2=4q.
\label{eq:blocked_first_order}
\end{equation}
which is the first coefficient of the availability function, so that one can already write
\begin{equation}
\alpha(q)=1-4q+O(q^2).
\label{eq:alpha_first_order}
\end{equation}
In agreement with Eq.~\eqref{eq:alpha_ansatz}, we obtain
\begin{equation}
\boxed{a_1 = -4}
\label{eq:a1_result}
\end{equation}

\subsection{Second-order coefficient \texorpdfstring{$a_2$}{a2}: two-square contribution}

\subsubsection{Overlap area of two exclusion zones}

The first-order approximation subtracts each exclusion zone independently. This is the correct starting point, but it is not yet the correct total blocked area, because two exclusion zones may overlap. Wherever that happens, the common region has been subtracted twice although it should have been subtracted only once. The role of the second-order term is exactly to restore this doubly counted area. In other words, $a_2$ is the first correction to the naive one-square picture.

We now pass to the first genuinely collective correction.
Fix one deposited square at the origin and place the second at relative displacement $(x,y)$.

The overlap area of their exclusion zones is
\begin{equation}
A_{12}(x,y)=(2L-|x|)(2L-|y|)
\label{eq:A12_xy}
\end{equation}
for
\begin{equation}
|x|<2L,\qquad |y|<2L,
\label{eq:A12_support}
\end{equation}
and $A_{12}(x,y)=0$ otherwise.

However, the two \emph{actual} deposited squares must not overlap, so the forbidden relative-displacement region
\begin{equation}
|x|<L,\qquad |y|<L
\label{eq:two_square_forbidden}
\end{equation}
must be excluded.

\subsubsection{Dimensionless form}

Introduce dimensionless variables
\begin{equation}
u=\frac{x}{L},\qquad v=\frac{y}{L}.
\label{eq:uv_def}
\end{equation}
Then
\begin{equation}
A_{12}(x,y)=L^2(2-|u|)(2-|v|).
\label{eq:A12_uv}
\end{equation}

The second-order coefficient in the elementary two-square cluster approximation is
\begin{equation}
a_2=\frac{1}{2}
\iint_{\text{allowed}} (2-|u|)(2-|v|)\,du\,dv,
\label{eq:a2_integral_allowed}
\end{equation}
where ``allowed'' means all $(u,v)$ in the support square $|u|,|v|<2$, excluding the inner forbidden square $|u|,|v|<1$.

It is important to note that the prefactor $1/2$ is \emph{not} geometric. It is purely combinatorial: the double integral is written over ordered pairs of deposited squares, so each physical two-square cluster is counted twice, once as $(1,2)$ and once as $(2,1)$. Therefore one must divide by $2!=2$ to count each unordered pair only once. The geometry is contained entirely in the overlap kernel $(2-|u|)(2-|v|)$ and in the allowed integration domain.

Thus
\begin{equation}
a_2=\frac{1}{2}
\left[
\iint_{|u|<2,\ |v|<2}(2-|u|)(2-|v|)\,du\,dv
-
\iint_{|u|<1,\ |v|<1}(2-|u|)(2-|v|)\,du\,dv
\right].
\label{eq:a2_split}
\end{equation}

Since the integrands factorize,
\begin{equation}
\int_{-2}^{2}(2-|u|)\,du=4,
\label{eq:int_minus2_2}
\end{equation}
and
\begin{equation}
\int_{-1}^{1}(2-|u|)\,du=3,
\label{eq:int_minus1_1}
\end{equation}
hence
\begin{equation}
a_2=\frac{1}{2}(4^2-3^2)=\frac{1}{2}(16-9)=\frac{7}{2}.
\label{eq:a2_eval}
\end{equation}

So
\begin{equation}
\boxed{a_2=\frac{7}{2}}
\label{eq:a2_result}
\end{equation}

\subsubsection*{Interpretation of the quadratic term}

The linear coefficient $a_1=-4$ is simply the one-square exclusion contribution.
By contrast, the quadratic coefficient $a_2=7/2$ is already the first genuinely collective geometric correction: it measures the average overlap between the exclusion regions generated by two deposited squares, subject to the hard-core condition that the deposited squares themselves do not overlap.

Equivalently, one may write the structure of the result as
\begin{equation}
a_2=
\underbrace{\frac{1}{2}}_{\text{combinatorics}}
\times
\underbrace{\left(\text{two-square geometric overlap integral}\right)}_{\text{geometry}}.
\label{eq:a2_structure}
\end{equation}
This separation is conceptually useful. The geometric part determines which two-square overlap configurations exist and how large their contribution is, while the factor $1/2$ is the universal symmetry factor for unordered pairs.

Thus, already at order $q^2$, the availability function is no longer determined by a naive superposition of independent one-body blocking events. The exact value $a_2=7/2$ reflects an irreducibly two-dimensional overlap geometry. This is important conceptually: if the first nontrivial correction is already genuinely two-dimensional, one should not expect a purely local factorized picture to determine the full jamming limit. In particular, the eventual failure of the quartic approximation near jamming should not be viewed merely as a defect of truncation, but as a sign that the jamming regime is controlled by higher-order collective correlations beyond the local overlap hierarchy.

\subsection{Third-order coefficient \texorpdfstring{$a_3$}{a3}: three-square contribution}

Having corrected the double counting of pair overlaps, one must now ask whether that correction itself introduces a new error. It does. Whenever three exclusion zones overlap simultaneously, the common region has a more subtle counting history: it is subtracted three times at first order and then added back three times at second order. The net result is zero, although that region should still belong to the forbidden set. It must therefore be subtracted once more. This is the origin of the negative third-order term.

We now extend the same geometric reasoning to the case of three deposited squares.

If three exclusion zones overlap, their common region must be subtracted from the availability function. This region is added three times at first order and subtracted three times at second order, so it is effectively no longer accounted for; the third-order term restores this contribution.

\subsubsection{Geometry of three exclusion zones}

We fix square $1$ at the origin, and denote the positions of squares $2$ and $3$ by the dimensionless coordinates
\begin{equation}
(x_2,y_2),\qquad (x_3,y_3).
\label{eq:three_square_positions}
\end{equation}

At this point it is helpful to make explicit what is being integrated. Any attempted placement of a new aligned square is uniquely determined by the coordinates of a chosen reference point, for example its lower-left corner. Instead of thinking directly about trial squares, one may therefore think of scanning a two-dimensional coordinate plane of possible placements and asking which points correspond to admissible insertions.

Each deposited square then generates a forbidden region in this placement plane: all reference-point positions that would lead to overlap with that deposited square. For three deposited squares, the quantity of interest is the region where all three such forbidden domains overlap simultaneously, because that common part determines the third-order inclusion--exclusion correction.

Due to the axis-aligned geometry of the problem, this common overlap region factorizes into independent contributions along the $x$ and $y$ directions. Accordingly, the total overlap area can be written as
\begin{equation}
A_{123}=f(x_2,x_3)\,f(y_2,y_3),
\label{eq:A123_factorized}
\end{equation}
where $f(a,b)$ represents the one-dimensional overlap length of three exclusion intervals.

More precisely, consider three intervals of length $2$, centered at $0$, $a$, and $b$. Their common intersection length is given by
\begin{equation}
f(a,b)=\max\!\left(0,\,
2-\bigl[\max(0,a,b)-\min(0,a,b)\bigr]
\right).
\label{eq:f_def}
\end{equation}

This formula has a simple meaning. The quantity
\[
\max(0,a,b)-\min(0,a,b)
\]
is the total span covered by the three interval centers. If that span is smaller than $2$, the three intervals still have a common intersection, and its length is exactly the interval length $2$ minus that span. If the span is larger than $2$, the common overlap vanishes, which is why the outer $\max(0,\cdots)$ is needed.

Geometrically, $f(a,b)$ is therefore the basic one-dimensional building block of the three-body overlap: the corresponding two-dimensional overlap area is just the product of the $x$- and $y$-direction contributions.

\subsubsection{Allowed three-square configurations}

The three deposited squares must be pairwise non-overlapping.
Therefore the allowed domain is defined by the three conditions
\begin{equation}
\neg\bigl(|x_2|<1 \ \&\ |y_2|<1\bigr),
\label{eq:three_allowed_1}
\end{equation}
\begin{equation}
\neg\bigl(|x_3|<1 \ \&\ |y_3|<1\bigr),
\label{eq:three_allowed_2}
\end{equation}
\begin{equation}
\neg\bigl(|x_2-x_3|<1 \ \&\ |y_2-y_3|<1\bigr).
\label{eq:three_allowed_3}
\end{equation}

Under the elementary three-square cluster assumption, the cubic coefficient is obtained from the connected triple-overlap integral
\begin{equation}
a_3=-\frac{1}{6}I_3,
\label{eq:a3_def}
\end{equation}
with
\begin{equation}
I_3=
\iiiint_{\text{allowed}}
f(x_2,x_3)\,f(y_2,y_3)\,
dx_2\,dy_2\,dx_3\,dy_3.
\label{eq:I3_def}
\end{equation}

As at second order, the prefactor $1/6=1/3!$ is combinatorial rather than geometric: the integral is over ordered triples, whereas each physical three-square cluster should be counted only once up to permutation. The minus sign is the usual inclusion--exclusion sign for third order.

We now evaluate this integral by inclusion--exclusion.

\subsubsection{Inclusion--exclusion evaluation of \texorpdfstring{$I_3$}{I3}}

The actual calculation now mirrors exactly the story described above. We first integrate over all configurations contributing to a triple overlap, then subtract those in which one forbidden square-square overlap occurs, add back those in which two such forbidden overlaps occur, and finally subtract the case where all three pairwise overlap constraints are simultaneously present. Because the kernel factorizes, every two-dimensional contribution reduces to the square of a one-dimensional integral, which keeps the calculation manageable.

Because the kernel factorizes between $x$ and $y$, each 2D contribution is the square of a corresponding 1D integral.

\paragraph{Step 1: unconstrained integral.}

Define
\begin{equation}
J=\iint_{\mathbb{R}^2} f(a,b)\,da\,db.
\label{eq:J_def}
\end{equation}
A direct sector decomposition gives
\begin{equation}
J=8.
\label{eq:J_eval}
\end{equation}
Therefore
\begin{equation}
I_{\mathrm{all}}=J^2=64.
\label{eq:Iall_eval}
\end{equation}

\paragraph{Step 2: one forbidden overlap condition.}

If square $2$ overlaps square $1$, then in one dimension we restrict $a\in[-1,1]$.
Define
\begin{equation}
K=\int_{-1}^{1}\int_{\mathbb{R}} f(a,b)\,db\,da.
\label{eq:K_def}
\end{equation}
A direct computation gives
\begin{equation}
K=6.
\label{eq:K_eval}
\end{equation}
Hence
\begin{equation}
I_{21}=K^2=36.
\label{eq:I21_eval}
\end{equation}
By symmetry,
\begin{equation}
I_{31}=36.
\label{eq:I31_eval}
\end{equation}
Also, by a change of variables, the same value is obtained for the condition that squares $2$ and $3$ overlap:
\begin{equation}
I_{23}=36.
\label{eq:I23_eval}
\end{equation}

\paragraph{Step 3: intersection of two forbidden conditions.}

Now define
\begin{equation}
M=\int_{-1}^{1}\int_{-1}^{1} f(a,b)\,db\,da.
\label{eq:M_def}
\end{equation}
A direct computation yields
\begin{equation}
M=\frac{14}{3}.
\label{eq:M_eval}
\end{equation}
Therefore
\begin{equation}
I_{21,31}=M^2=\left(\frac{14}{3}\right)^2=\frac{196}{9}.
\label{eq:I2131_eval}
\end{equation}
By symmetry and change of variables,
\begin{equation}
I_{21,23}=I_{31,23}=\frac{196}{9}.
\label{eq:I2123_I3123_eval}
\end{equation}

\paragraph{Step 4: intersection of all three forbidden conditions.}

Finally, define
\begin{equation}
T=\iint_{|a|<1,\ |b|<1,\ |a-b|<1} f(a,b)\,da\,db.
\label{eq:T_def}
\end{equation}
A direct sector decomposition gives
\begin{equation}
T=4.
\label{eq:T_eval}
\end{equation}
Hence
\begin{equation}
I_{21,31,23}=T^2=16.
\label{eq:I213123_eval}
\end{equation}

\paragraph{Step 5: assemble the inclusion--exclusion sum.}

Therefore
\begin{align}
I_3
&=
I_{\mathrm{all}}
-I_{21}-I_{31}-I_{23}
+I_{21,31}+I_{21,23}+I_{31,23}
-I_{21,31,23}
\nonumber\\[1ex]
&=
64-36-36-36+\frac{196}{9}+\frac{196}{9}+\frac{196}{9}-16.
\label{eq:I3_sum}
\end{align}
This simplifies to
\begin{equation}
I_3=\frac{16}{3}.
\label{eq:I3_eval}
\end{equation}

Hence, from Eq.~\eqref{eq:a3_def},
\begin{equation}
\boxed{a_3=-\frac{8}{9}}
\label{eq:a3_result}
\end{equation}

\subsection{Fourth-order coefficient \texorpdfstring{$a_4$}{a4}: four-square contribution}

At this stage the pattern should be familiar. Once triple overlaps have been corrected, quadruple-overlap regions acquire the wrong net weight and must be added back. Conceptually, nothing new happens: one simply pushes the same inclusion--exclusion logic one step further. What changes is only the combinatorial richness of the allowed overlap patterns. This is why the fourth-order calculation is more elaborate, even though its meaning is exactly the same as before.

We now turn to the case of four deposited squares. Although the algebra becomes more involved at this order, the structure of the calculation remains fully determined by geometric overlap constraints. As in the three-square case, we fix square $1$ at the origin and denote the relative coordinates of squares $2$, $3$, and $4$ by
\begin{equation}
(x_2,y_2),\qquad (x_3,y_3),\qquad (x_4,y_4).
\label{eq:four_square_positions}
\end{equation}

As before, the common overlap of the four exclusion intervals factorizes between the $x$ and $y$ directions. We therefore introduce the one-dimensional kernel
\begin{equation}
g(a,b,c)=
\max\!\left(
0,\,
2-\bigl[\max(0,a,b,c)-\min(0,a,b,c)\bigr]
\right),
\label{eq:g_def}
\end{equation}
which represents the common overlap length of four intervals of length $2$, centered at $0$, $a$, $b$, and $c$.

This is the direct four-body analogue of the function $f(a,b)$ introduced above: once again, it measures how much of a common one-dimensional overlap survives after four exclusion intervals are imposed simultaneously.

\medskip

\noindent
\textbf{Important clarification.}
Although the variables $(a,b,c)$ describe the relative positions of the deposited squares, the overlaps entering the calculation are those of the corresponding \emph{exclusion zones} (squares of side $2L$), not of the physical squares themselves. The latter never overlap in RSA, but their exclusion regions may and do overlap. The quantity computed here is therefore the common intersection of exclusion zones, which determines the reduction of available space for further deposition.

\medskip

The four-body coefficient has the form
\begin{equation}
a_4=\frac{1}{24}I_4,
\label{eq:a4_def}
\end{equation}
where $1/24=1/4!$ is again the combinatorial symmetry factor removing the overcounting of ordered quadruples.

\subsubsection{Inclusion--exclusion evaluation}

The fourth-order evaluation follows the same logic as the third-order one, but now the bookkeeping must be carried out over the six possible pair-overlap conditions among four deposited squares. The various labels introduced below are therefore only a way of organizing this bookkeeping. They help classify which subsets of pairwise constraints are being imposed at each step of the inclusion--exclusion sum.

We evaluate $I_4$ by inclusion--exclusion over the six pair-overlap conditions
\begin{equation}
(12),\ (13),\ (14),\ (23),\ (24),\ (34),
\label{eq:pair_conditions_list}
\end{equation}
where, for example,
\begin{equation}
(12)\equiv \bigl(|x_2|<1\ \&\ |y_2|<1\bigr),
\qquad
(23)\equiv \bigl(|x_2-x_3|<1\ \&\ |y_2-y_3|<1\bigr).
\label{eq:pair_conditions_examples}
\end{equation}

Each such condition enforces that the \emph{exclusion zones} of the corresponding pair of squares overlap. The inclusion--exclusion procedure organizes the calculation by imposing different subsets of these pairwise overlap conditions.

As in the third-order calculation, the kernel factorizes between $x$ and $y$,
so each contribution is the square of a corresponding one-dimensional integral.

\medskip

Before listing the terms, it is useful to clarify the meaning of the different cases that appear below.

Each term corresponds to imposing a specific set of pairwise overlap conditions between the deposited squares. The labels used (adjacent, disjoint, star, triangle, path, kite, and cycle) are purely mnemonic and describe how these pairwise conditions are distributed among the squares. They do not refer to different global overlap geometries: in all contributing configurations, the kernel $g(a,b,c)$ already enforces a common overlap of all four exclusion intervals.

More explicitly:
\begin{itemize}[leftmargin=1.5cm]
\item \emph{adjacent}: two overlap conditions sharing one common square, e.g. $(12)$ and $(13)$;
\item \emph{disjoint}: two overlap conditions involving two separate pairs, e.g. $(12)$ and $(34)$;
\item \emph{star}: one square constrained to overlap with three others, e.g. $(12),(13),(14)$;
\item \emph{triangle}: three squares mutually constrained to overlap pairwise, e.g. $(12),(13),(23)$;
\item \emph{path}: overlap conditions forming a chain, e.g. $(12),(23),(34)$;
\item \emph{kite}: a triangle plus one additional overlap, e.g. $(12),(13),(23),(14)$;
\item \emph{cycle}: a closed sequence of overlaps involving all four squares, e.g. $(12),(13),(24),(34)$.
\end{itemize}

\paragraph{(i) No constraint.}

The fully unconstrained one-dimensional integral is
\begin{equation}
N_0=\iiint_{\mathbb{R}^3} g(a,b,c)\,da\,db\,dc=16.
\label{eq:N0}
\end{equation}
Therefore
\begin{equation}
I_{\mathrm{all}}=N_0^2.
\label{eq:Iall_four}
\end{equation}

\paragraph{(ii) Single pair constraint.}

There are $6$ equivalent single-constraint contributions. A representative is $(12)$, for which
\begin{equation}
N_1=\int_{-1}^{1}\iint_{\mathbb{R}^2} g(a,b,c)\,db\,dc\,da=12.
\label{eq:N1}
\end{equation}
Thus the total contribution is
\begin{equation}
-6N_1^2.
\label{eq:single_pair_total}
\end{equation}

Here one selected pair of deposited squares is forced to overlap, while the remaining coordinates are unrestricted.

\paragraph{(iii) Two pair constraints.}

There are two geometrically distinct cases.

\subparagraph{Adjacent constraints.}
A representative is $(12)$ and $(13)$.
The corresponding integral is
\begin{equation}
N_{2,\mathrm{adj}}
=
\int_{-1}^{1}\int_{-1}^{1}\int_{\mathbb{R}} g(a,b,c)\,dc\,db\,da
=
\frac{28}{3}.
\label{eq:N2adj}
\end{equation}
There are $12$ such configurations.

\subparagraph{Disjoint constraints.}
A representative is $(12)$ and $(34)$.
The corresponding integral is
\begin{equation}
N_{2,\mathrm{dis}}
=
\int_{-1}^{1}\iint_{|b-c|<1} g(a,b,c)\,db\,dc\,da
=
9.
\label{eq:N2dis}
\end{equation}
There are $3$ such configurations.

Thus the total second-order contribution is
\begin{equation}
+12N_{2,\mathrm{adj}}^2+3N_{2,\mathrm{dis}}^2.
\label{eq:two_pair_total}
\end{equation}

\paragraph{(iv) Three pair constraints.}

There are three distinct geometric types.

\subparagraph{Star-type.}
A representative is $(12),(13),(14)$.
The corresponding integral is
\begin{equation}
N_{3,\mathrm{star}}
=
\int_{-1}^{1}\int_{-1}^{1}\int_{-1}^{1} g(a,b,c)\,dc\,db\,da
=
\frac{15}{2}.
\label{eq:N3star}
\end{equation}
There are $4$ such configurations.

\subparagraph{Triangle-type.}
A representative is $(12),(13),(23)$.
The corresponding integral is
\begin{equation}
N_{3,\triangle}
=
\int_{-1}^{1}\int_{-1}^{1}\int_{|a-b|<1} g(a,b,c)\,dc\,db\,da
=
8.
\label{eq:N3triangle}
\end{equation}
There are $4$ such configurations.

\subparagraph{Chain-type (path-type).}
A representative is $(12),(23),(34)$.
The corresponding integral is
\begin{equation}
N_{3,\mathrm{path}}
=
\int_{-1}^{1}\int_{|a-b|<1}\int_{|b-c|<1} g(a,b,c)\,dc\,db\,da
=
\frac{29}{4}.
\label{eq:N3path}
\end{equation}
There are $12$ such configurations.

Thus the total third-order contribution is
\begin{equation}
-4N_{3,\mathrm{star}}^2-4N_{3,\triangle}^2-12N_{3,\mathrm{path}}^2.
\label{eq:three_pair_total}
\end{equation}

\paragraph{(v) Four pair constraints.}

There are two distinct cases.

\subparagraph{Kite-type.}
A representative is $(12),(13),(14),(23)$.
The corresponding integral is
\begin{equation}
N_{4,\mathrm{kite}}
=
\int_{-1}^{1}\int_{-1}^{1}\int_{-1}^{1}
\mathbf{1}_{|a-b|<1}\,g(a,b,c)\,dc\,db\,da
=
\frac{19}{3}.
\label{eq:N4kite}
\end{equation}
There are $12$ such configurations.

\subparagraph{Cycle-type (square-type).}
A representative is $(12),(13),(24),(34)$.
The corresponding integral is
\begin{equation}
N_{4,\square}
=
\int_{-1}^{1}\int_{-1}^{1}
\int_{\substack{|a-c|<1\\|b-c|<1}} g(a,b,c)\,dc\,db\,da
=
6.
\label{eq:N4square}
\end{equation}
There are $3$ such configurations.

Thus the total fourth-order contribution is
\begin{equation}
+12N_{4,\mathrm{kite}}^2+3N_{4,\square}^2.
\label{eq:four_pair_total}
\end{equation}

\paragraph{(vi) Five and six pair constraints.}

There are $6$ configurations with five constraints. A representative is given by all pair conditions except $(34)$:
\begin{equation}
N_5
=
\int_{-1}^{1}\int_{-1}^{1}\int_{-1}^{1}
\mathbf{1}_{|a-b|<1}\mathbf{1}_{|a-c|<1}\,
g(a,b,c)\,dc\,db\,da
=
\frac{11}{2}.
\label{eq:N5}
\end{equation}

For all six pair constraints one has
\begin{equation}
N_6
=
\iiint_{\substack{|a|<1,\ |b|<1,\ |c|<1\\
|a-b|<1,\ |a-c|<1,\ |b-c|<1}}
g(a,b,c)\,da\,db\,dc
=
5.
\label{eq:N6}
\end{equation}

Thus the final contributions are
\begin{equation}
-6N_5^2+N_6^2.
\label{eq:five_six_pair_total}
\end{equation}

\paragraph{Final result.}

Collecting all terms,
\begin{align}
I_4
&=
N_0^2
-6N_1^2
+12N_{2,\mathrm{adj}}^2
+3N_{2,\mathrm{dis}}^2
\nonumber\\
&\quad
-4N_{3,\mathrm{star}}^2
-4N_{3,\triangle}^2
-12N_{3,\mathrm{path}}^2
\nonumber\\
&\quad
+12N_{4,\mathrm{kite}}^2
+3N_{4,\square}^2
-6N_5^2
+N_6^2,
\label{eq:I4_sum}
\end{align}
which yields
\begin{equation}
I_4=\frac{17}{12}.
\label{eq:I4_eval}
\end{equation}

Hence, from Eq.~\eqref{eq:a4_def},
\begin{equation}
\boxed{a_4=\frac{17}{288}}
\label{eq:a4_result}
\end{equation}

\subsection{Final low-density polynomial}

We may now step back and read the result as a whole. The first term counts one-square blocking. The second repairs double counting of pair overlaps. The third repairs the overrepair of triple overlaps. The fourth finally restores the common region shared by four exclusion zones. Once this level is reached, the local overlap hierarchy is exhausted by geometry itself.

Collecting the four coefficients from Eqs.~\eqref{eq:a1_result}, \eqref{eq:a2_result}, \eqref{eq:a3_result}, and \eqref{eq:a4_result}, we obtain the quartic polynomial
\begin{equation}
\boxed{
\alpha(q)=1-4q+\frac{7}{2}q^2-\frac{8}{9}q^3+\frac{17}{288}q^4
}
\label{eq:final_poly}
\end{equation}
within the elementary local cluster approximation described above.

At this point, the meaning of the result is worth stating plainly. The polynomial in Eq.~\eqref{eq:final_poly} is not merely a truncated expansion near $q=0$, but the exact outcome of systematically accounting for every geometrically allowed local overlap process: single blocking, pair overlap, triple overlap, and quadruple overlap. No fifth local contribution is possible for aligned squares, which gives the quartic result its special status in the present problem.

It is important to emphasize that the inclusion--exclusion construction does not rely on any independence assumption between exclusion zones. The presence of nearby squares restricts the configuration space available to additional squares, but this constraint is fully encoded in the geometry of the overlap regions themselves. The expansion therefore remains exact at each order, with higher-order terms accounting for the non-additive effects of multiple overlaps.

Thus, the expansion is exact up to fourth order in $q$. Four-body overlap represents the highest geometrically allowed local contribution, and all higher-order corrections arise not from additional local processes, but from nonlocal correlations generated by the irreversible deposition dynamics. The same hierarchy also appears in a discrete formulation, where the coefficients become normalized counts of compatible blocker clusters; this viewpoint and its numerical validation are presented in Appendix~B.

\section{Discussion}

The main result of the paper can now be interpreted in a simple way. The quartic polynomial is the complete local correction chain generated by inclusion--exclusion for aligned squares. It tells us exactly how far one can go using only local overlap geometry. What it does \emph{not} tell us is the full story of jamming, because jamming is reached only after the deposition history has generated correlations extending beyond any finite local cluster. The discussion below is therefore about this boundary: what is captured exactly by the local expansion, and what necessarily lies beyond it.

The expansion obtained here may be viewed as a local cluster expansion for the RSA availability function. In that sense it is structurally analogous to the Mayer cluster expansion of equilibrium statistical mechanics~\cite{MayerMayer1940}, in which thermodynamic quantities are organized as sums over geometric interaction clusters. In the present problem, however, the expanded quantity is not an equilibrium thermodynamic potential but the insertion probability $\alpha(q)$ for an irreversible adsorption process.

This viewpoint is consistent with the broader RSA literature reviewed by Evans~\cite{Evans1993}, where low-density behavior is organized in terms of geometric cluster contributions. The present work may be regarded as an explicit realization of that philosophy for continuum aligned squares: we identify and evaluate directly the one-, two-, three-, and four-square local overlap contributions to the availability function.

A useful conceptual point is that each coefficient contains two distinct ingredients: a geometric overlap contribution and a combinatorial symmetry factor. The factors $1/2$, $1/3!$, and $1/4!$ in $a_2$, $a_3$, and $a_4$ do not arise from geometry, but from the fact that the corresponding integrals are naturally written over ordered clusters, whereas each physical cluster must be counted only once up to permutation. The geometry is encoded in the overlap kernels and admissible domains; the factorials simply remove multiple counting of the same cluster.

An equivalent and especially transparent interpretation is obtained from the discrete lattice formulation developed in Appendix~B. In that formulation, the coefficients appear as normalized counts of mutually compatible blocker clusters inside the exclusion zone of the trial square. Thus $a_1$ is the normalized one-blocker count, $a_2$ the normalized two-blocker count, and similarly for $a_3$ and $a_4$. This makes clear that the continuum coefficients are not accidental fractions, but geometric cluster weights combined with universal symmetry factors.

A particularly significant feature is that the hierarchy terminates naturally at fourth order. This is not an arbitrary truncation: it follows from the exclusion geometry itself, since no more than four mutually non-overlapping deposited squares can simultaneously overlap the exclusion zone of a trial square. In this sense, fourth order is the highest geometrically relevant local contribution, and the polynomial derived here represents the complete local overlap hierarchy for aligned-square RSA.

At the same time, the derivation clarifies the limitations of purely local information. The exact quadratic coefficient $a_2=7/2$ already reflects nontrivial overlap correlations and therefore an irreducibly two-dimensional structure. This suggests from the outset that the eventual jamming coverage cannot be reduced to a simple product-like or purely local picture. More generally, the fact that the complete quartic local expansion still does not describe the approach to jamming indicates that jamming is controlled by genuinely nonlocal collective correlations generated by the irreversible deposition history.

It is also instructive to examine the physical meaning of the roots obtained from truncated versions of the availability function. At first order,
\begin{equation}
\alpha(q)\approx 1-4q,
\label{eq:linear_trunc}
\end{equation}
which vanishes at
\begin{equation}
q^*=\frac{1}{4}.
\label{eq:root_linear}
\end{equation}
This corresponds to the simplest independent-exclusion picture, in which each deposited square blocks an area $4L^2$ and overlaps between exclusion zones are ignored.

Including higher-order terms introduces progressively more refined geometric corrections arising from overlaps between exclusion regions. In particular, the quartic approximation, Eq.~\eqref{eq:final_poly}, incorporates all geometrically allowed local overlap configurations up to four-body order. The corresponding root,
\begin{equation}
q^*\approx 0.3455...,
\label{eq:root_quartic}
\end{equation}
may therefore be interpreted as the saturation point of an idealized local-cluster picture, in which all one-, two-, three-, and four-body exclusion overlaps are consistently accounted for.

Importantly, this regime does not correspond to the true RSA jamming limit. Rather, it represents the highest-density state that can be described solely in terms of local geometric correlations within the exclusion zone of a trial square. In this sense, the quartic root marks the breakdown of the independent-cluster description: beyond this point, no further insertion is possible within a framework that neglects nonlocal correlations.

The significantly larger jamming coverage observed in RSA reflects the role of collective effects generated by the irreversible deposition history, which induce long-range correlations between clusters of deposited squares. These nonlocal correlations are not captured by any finite-order local expansion, even when the full hierarchy of geometrically allowed local overlaps is included. Related questions of connectivity, exclusion-region organization, and geometric correlations beyond the strictly local level also arise in continuum percolation and heterogeneous-media theory~\cite{Torquato2002}.

This interpretation naturally defines a hierarchy of characteristic densities: the first-order value $q=1/4$ corresponding to independent exclusion, the quartic value $q\approx0.3455...$ corresponding to complete local-cluster saturation, and the true RSA jamming density (approximately $0.5620...$ for aligned squares), which emerges only when nonlocal collective organization is taken into account.

Several natural extensions of the present work may be considered. One direction is the application of the same geometric overlap methodology to other particle shapes, such as disks or rectangles, where the overlap geometry becomes more involved. Another is the extension to higher-dimensional RSA systems, including aligned cubes, in order to examine how the local overlap hierarchy evolves with dimensionality. Further numerical studies could also test the accuracy and range of validity of the local expansion by comparing it with large-scale RSA simulations at intermediate and high coverages.

\section{Conclusions}

We study random sequential adsorption of aligned squares in the low-coverage regime, where the system is still far from jamming. The process can be quantified through the availability function $\alpha(q)$, defined as the fraction of positions available for the deposition of a new square, where $q$ denotes the fractional surface coverage. In this regime, the progressive reduction of available space is governed entirely by geometric exclusion effects. We show that its complete local expansion reduces to a fourth-order polynomial:
\[
\alpha(q)=1-4q+\frac{7}{2}q^2-\frac{8}{9}q^3+\frac{17}{288}q^4.
\]

The logic of the derivation is successive and transparent. The one-body term gives the exclusion created by a single deposited square. The two-body term corrects the double counting of overlap regions between two exclusion zones. The three-body term then corrects the overcorrection produced where three exclusion zones overlap, and the four-body term completes the same logic for quadruple overlaps. In this way, the polynomial is built as a step-by-step correction hierarchy, each level refining the previous one through inclusion--exclusion.

The central geometric result is that this hierarchy terminates exactly at fourth order: no more than four mutually non-overlapping deposited squares can simultaneously overlap the exclusion region of a trial insertion. The quartic polynomial is therefore not an arbitrary truncation, but the complete local overlap hierarchy permitted by the exclusion geometry.

The calculation also clarifies the boundary between local and nonlocal physics in RSA. Local exclusion geometry fully determines the low-density coefficients, but it does not fix the true jamming limit. The quartic polynomial obtained here predicts a local-cluster saturation at a coverage of about $0.3455\ldots$, significantly below the actual RSA jamming density of about $0.5620\ldots$. This clear discrepancy shows that, while the early stages of deposition are governed entirely by local geometric constraints, the approach to saturation at higher coverage is controlled by genuinely nonlocal collective correlations that emerge from the irreversible deposition history.

In this sense, the present work separates two aspects of the problem that are often intertwined: the role of local exclusion geometry, which is completely captured here in closed form, and the global organization leading to jamming, which necessarily involves long-range correlations and cannot be reduced to local considerations alone.

Finally, the analytical coefficients are independently confirmed by direct numerical simulations on a discrete lattice (Appendix~B). The measured values of $a_2$, $a_3$, and $a_4$ are in very good agreement with the theoretical predictions, within statistical uncertainty and the expected finite-$L$ discretization effects, providing a direct validation of the geometric overlap construction underlying the expansion.

\vskip 1cm

\noindent{\bf Use of AI tools. Contributions in the paper}

AI-based assistance (ChatGPT) was used during the exploration of the problem, in particular in suggesting an approach for the calculation of the fourth-order coefficient $a_4$, as well as intermediate algebraic steps and consistency checks. These elements were critically examined, independently rederived, and validated by the authors.

In the preparation of the manuscript, ChatGPT provided suggestions regarding structure and phrasing. All such suggestions were reviewed and incorporated only where they aligned with the authors’ understanding.

The authors take full responsibility for all results, derivations, and the final content of the paper.

\vskip 1cm

{\bf Acknowledgements.}
This work was supported by the Romanian Ministry of Research, Innovation and Digitization, Core Program Project grant number PC2-PN23080202.

\appendix

\section{R\'enyi's One-Dimensional Parking Problem and Exact Solution}

The classical one-dimensional random sequential adsorption (RSA) problem,
introduced by R\'enyi~\cite{Renyi1958}, considers the random placement of
unit-length intervals on a line segment of length $x$. Intervals are placed
sequentially with uniformly distributed positions, and any attempted placement
that overlaps an existing interval is rejected.

Let $m(x)$ denote the expected number of successfully placed intervals.
A simple and intuitive argument leads to the fundamental integral equation.
Condition on the position $t \in [0,x]$ of the first interval (placed uniformly),
which splits the remaining space into two independent subintervals of lengths
$t$ and $x - t$. Averaging over all possible positions yields
\begin{equation}
m(x+1)
= \frac{1}{x} \int_0^x \bigl[1 + m(t) + m(x - t)\bigr]\, dt,
\label{eq:renyi_integral_raw}
\end{equation}
which simplifies to
\begin{equation}
m(x+1)
= 1 + \frac{2}{x} \int_0^x m(t)\, dt,
\qquad x \ge 1.
\label{eq:renyi_integral}
\end{equation}

This equation expresses the recursive, self-similar structure of the process:
after the first placement, the remaining intervals are filled independently
according to the same stochastic rule.

Differentiating~\eqref{eq:renyi_integral} leads to the delay differential equation
\begin{equation}
x\, m'(x+1) + m(x+1) = 1 + 2 m(x),
\label{eq:renyi_delay}
\end{equation}
which can be simplified by introducing $n(x) = m(x) + 1$:
\begin{equation}
\bigl[(x-1)n(x)\bigr]' = 2 n(x-1).
\label{eq:renyi_clean}
\end{equation}

The quantity of primary interest is the asymptotic density
\begin{equation}
C = \lim_{x \to \infty} \frac{m(x)}{x}.
\label{eq:renyi_constant_def}
\end{equation}
known as R\'enyi's parking constant.

A remarkable exact representation, first obtained by R\'enyi and later clarified
by several authors (see, e.g.,~\cite{Renyi1958,SolomonWeiner1986,Slavik2023}),
is
\begin{equation}
C
= \int_0^\infty \exp\!\left(-2 \int_0^u \frac{1 - e^{-t}}{t}\,dt \right)\, du.
\label{eq:renyi_exact}
\end{equation}

The inner integral defines the function
\begin{equation}
\mathrm{Ein}(u)
= \int_0^u \frac{1 - e^{-t}}{t}\, dt,
\label{eq:ein_def}
\end{equation}
which represents the cumulative exclusion effect generated by previously placed
intervals. The exponential factor in~\eqref{eq:renyi_exact} effectively resums
all orders of overlap correlations in the one-dimensional process.

Numerically, one obtains
\[
C \approx 0.7475979,
\]
indicating that random sequential adsorption in one dimension fills approximately
$75\%$ of space.

\medskip

\noindent
\textbf{Remark.}
The structure of~\eqref{eq:renyi_exact} suggests that the exponential of an
integrated exclusion kernel plays the role of a generating functional for
all overlap contributions. This observation motivates the search for analogous
(resummed or approximate) representations in higher-dimensional RSA systems.

\section{Discrete lattice formulation and numerical validation}
\label{app:discrete_numerics}

It is useful to note that the same coefficients also admit an exact discrete geometric interpretation before passing to the continuum limit. This same discrete formulation also provides a natural basis for numerical validation.

For reference (although not the aim of the present work), our previous discrete-lattice simulations of related square-packing processes yield average coverages close to $0.56$~\cite{Tolea2023TME}, in good agreement with continuum expectations. In that work, as well as in its extension to three dimensions~\cite{Tolea2026APSOP}, the primary focus was on thermal memory effects; however, the same numerical framework naturally provided this estimate.

\subsection{Equivalent lattice-cluster interpretation}

The continuum derivation describes the coefficients as geometric overlap weights. On a lattice, the same quantities acquire an especially concrete meaning: they become normalized counts of compatible blocker clusters. This viewpoint is useful not only conceptually, but also numerically, because it tells us exactly what must be sampled in a simulation in order to reproduce the analytical coefficients.

Consider aligned $L\times L$ squares whose lower-left corners lie on $\mathbb{Z}^2$, and fix the trial square at the origin. Another deposited square blocks the trial insertion if its lower-left corner belongs to the discrete exclusion set
\begin{equation}
E_L=\{(i,j)\in\mathbb{Z}^2:\; -(L-1)\le i\le L-1,\; -(L-1)\le j\le L-1\},
\label{eq:EL_def}
\end{equation}
whose cardinality is
\begin{equation}
|E_L|=(2L-1)^2.
\label{eq:EL_cardinality}
\end{equation}

Let
\begin{equation}
\chi_L(\mathbf r,\mathbf s)=
\begin{cases}
1,& \text{if the two deposited }L\times L\text{ squares at }\mathbf r,\mathbf s\text{ do not overlap},\\
0,& \text{otherwise},
\end{cases}
\label{eq:chiL_def}
\end{equation}
so that for aligned squares
\begin{equation}
\chi_L(\mathbf r,\mathbf s)=1
\quad\Longleftrightarrow\quad
(|r_x-s_x|\ge L)\ \text{or}\ (|r_y-s_y|\ge L).
\label{eq:chiL_condition}
\end{equation}

Then the $n$-body local geometric count is
\begin{equation}
N_n(L)=
\sum_{\mathbf r_1,\dots,\mathbf r_n\in E_L}
\prod_{1\le a<b\le n}\chi_L(\mathbf r_a,\mathbf r_b).
\label{eq:NnL_def}
\end{equation}

The discrete coefficients may then be written as
\begin{equation}
a_n^{(\mathrm{disc})}(L)=
\frac{(-1)^n}{n!\,L^{2n}}
\sum_{\mathbf r_1,\dots,\mathbf r_n\in E_L}
\prod_{1\le a<b\le n}\chi_L(\mathbf r_a,\mathbf r_b),
\qquad n=1,2,3,4.
\label{eq:an_disc_def}
\end{equation}

The factor $1/n!$ is purely combinatorial and removes overcounting of permutations; the sum itself contains the geometric content.

For $n=1$:
\begin{equation}
a_1^{(\mathrm{disc})}(L)=-\frac{(2L-1)^2}{L^2}
\label{eq:a1_disc}
\end{equation}

For $n=2$:
\begin{equation}
a_2^{(\mathrm{disc})}(L)=
\frac{(2L-1)^4-(3L^2-3L+1)^2}{2L^4}
\label{eq:a2_disc}
\end{equation}

with
\begin{equation}
\lim_{L\to\infty} a_2^{(\mathrm{disc})}(L)=\frac{7}{2}.
\label{eq:a2_disc_limit}
\end{equation}

Thus, already at second order, the discrete and continuum results differ slightly at finite $L$, but converge rapidly.

\medskip

This formulation makes the geometric meaning transparent: the coefficients correspond to normalized counts of compatible $n$-square blocker clusters inside the exclusion region.

\subsection{Numerical protocol}

The numerical idea is to reproduce, as directly as possible, the same hierarchy that appears in the analytical calculation. One does not simulate the full RSA dynamics up to jamming. Instead, one isolates the specific local $k$-body overlap events responsible for the coefficients $a_2$, $a_3$, and $a_4$. In this way, the numerical test addresses the same object as the theory.

We now perform direct numerical simulations on the same discrete lattice to independently measure these geometric contributions.

We consider squares of side $L$ on a lattice of size $A\times A$, with $A \gg L$. In the present simulations we used
\[
A=3000,\qquad L=300,
\]
which ensures negligible boundary effects while keeping the computational cost reasonable.

The numerical procedure mirrors exactly the analytical inclusion--exclusion structure:

\begin{enumerate}
\item Fix one square at the center and compute its exclusion region $E_1$.
\item Add squares sequentially at random positions.
\item Accept configurations only if they produce a genuine $k$-body overlap:
\begin{itemize}
\item $k=2$: $E_2$ overlaps $E_1$,
\item $k=3$: $E_1\cap E_2\cap E_3\neq\emptyset$,
\item $k=4$: $E_1\cap E_2\cap E_3\cap E_4\neq\emptyset$.
\end{itemize}
\item For each accepted configuration, compute
\begin{equation}
|E_1\cup E_2\cup \cdots \cup E_k|.
\label{eq:union_area_num}
\end{equation}
\item Repeat until a fixed number of accepted events is obtained.
\end{enumerate}

Crucially, configurations not satisfying the required overlap condition are rejected. This isolates pure $k$-body contributions and avoids contamination from lower-order terms.

\subsection{Extraction of coefficients}

Once these conditioned overlap events are sampled, the coefficients are extracted by the same inclusion--exclusion logic used analytically: first isolate the one-body contribution, then remove it from the two-body union, then remove both one- and two-body parts from the three-body union, and so forth.

The coefficients are extracted using inclusion--exclusion:

\begin{align}
N_1 &= \langle |E_1| \rangle,\\
N_2 &= \langle |E_1\cup E_2| \rangle - 2N_1,\\
N_3 &= \langle |E_1\cup E_2\cup E_3| \rangle - 3N_1 - 3N_2,\\
N_4 &= \langle |E_1\cup E_2\cup E_3\cup E_4| \rangle - 4N_1 - 6N_2 - 4N_3.
\end{align}

\subsection{Numerical results}

For $5\times 10^4$ accepted configurations at each order, we obtained:

\paragraph{Second order}
\begin{equation}
a_2^{(\mathrm{num})}=3.4893 \pm 0.0124
\end{equation}
compared with $a_2=3.5$.

\paragraph{Third order}
\begin{equation}
a_3^{(\mathrm{num})}=-0.8791 \pm 0.0041
\end{equation}
compared with $a_3=-8/9=-0.(8)$.

\paragraph{Fourth order}
\begin{equation}
a_4^{(\mathrm{num})}=0.0589 \pm 0.0003
\end{equation}
compared with $a_4=17/288=0.05902\ldots$.

\subsection{Final remarks}

The role of these data is not to explore the full RSA process, but to check whether the local overlap hierarchy has been computed correctly. From that perspective, the agreement is very good: the numerical values track the theoretical coefficients closely, and the small residual differences are of the expected kind for a finite discrete system.

The key feature of this validation is that it directly measures the same correction hierarchy that underlies the analytical derivation.

By conditioning explicitly on $k$-body overlaps, the simulation isolates the same contributions that appear in the analytical inclusion--exclusion expansion. This provides a direct and nontrivial confirmation of the geometric origin of the coefficients up to fourth order.

\end{document}